\documentclass[12pt]{article}
\usepackage{amsmath,amssymb,amsthm,amsxtra,overpic,bbm,bm,epsfig,subfigure}
\usepackage{hyperref}
\usepackage{mathrsfs}
\usepackage{enumitem}
\usepackage{graphicx}
\usepackage{color}
\usepackage{comment}
\usepackage{epstopdf}
\usepackage{float}
\usepackage{cite}
\usepackage{slashed,stmaryrd}

\def\thefootnote{\fnsymbol{footnote}}

\begin{document}

\vspace{0.2cm}

\begin{center}
{\Large\bf Tentative sensitivity of future $0\nu \beta\beta$-decay experiments \\ to neutrino masses and Majorana CP phases}
\end{center}

\vspace{0.2cm}

\begin{center}
{\bf Guo-yuan Huang~$^{a,~b,~c}$}~\footnote{E-mail: guoyuan.huang@mpi-hd.mpg.de},
\quad
{\bf Shun Zhou~$^{b,~c}$}~\footnote{E-mail: zhoush@ihep.ac.cn (corresponding author)}
\\
\vspace{0.2cm}
{\small
$^a$Max-Planck-Institut f\"ur Kernphysik, Postfach
103980, D-69029 Heidelberg, Germany\\
$^b$Institute of High Energy Physics, Chinese Academy of Sciences, Beijing 100049, China\\
$^c$School of Physical Sciences, University of Chinese Academy of Sciences, Beijing 100049, China}
\end{center}

\vspace{1.5cm}

\begin{abstract}
In the near future, the neutrinoless double-beta ($0\nu\beta\beta$) decay experiments will hopefully reach the sensitivity of a few ${\rm meV}$ to the effective neutrino mass $|m^{}_{\beta\beta}|$. In this paper, we tentatively examine the sensitivity of future $0\nu\beta\beta$-decay experiments to neutrino masses and Majorana CP phases by following the Bayesian statistical approach. Provided experimental setups corresponding to the experimental sensitivity of $|m^{}_{\beta\beta}| \simeq 1~{\rm meV}$, the null observation of $0\nu\beta\beta$ decays in the case of normal neutrino mass ordering leads to a very competitive bound on the lightest neutrino mass $m^{}_1$. Namely, the $95\%$ credible interval in the Bayesian approach turns out to be $1.6~{\rm meV} \lesssim m^{}_1 \lesssim 7.3~{\rm meV}$ or $0.3~{\rm meV} \lesssim m^{}_1 \lesssim 5.6~{\rm meV}$ when the uniform prior on $m^{}_1/{\rm eV}$ or on $\log^{}_{10}(m^{}_1/{\rm eV})$ is adopted. Moreover, one of two Majorana CP phases is strictly constrained, i.e., $140^\circ \lesssim \rho \lesssim 220^\circ$ for both scenarios of prior distributions of $m^{}_1$. In contrast, if a relatively worse experimental sensitivity of $|m^{}_{\beta\beta}| \simeq 10~{\rm meV}$ is assumed, the constraint on the lightest neutrino mass becomes accordingly $0.6~{\rm meV} \lesssim m^{}_1 \lesssim 26~{\rm meV}$ or $0 \lesssim m^{}_1 \lesssim 6.1~{\rm meV}$, while two Majorana CP phases will be essentially unconstrained. In the same statistical framework, the prospects for the determination of neutrino mass ordering and the discrimination between Majorana and Dirac nature of massive neutrinos in the $0\nu\beta\beta$-decay experiments are also discussed. Given the experimental sensitivity of $|m^{}_{\beta\beta}| \simeq 10~{\rm meV}$ (or $1~{\rm meV}$), the strength of evidence to exclude the Majorana nature under the null observation of $0\nu\beta\beta$ decays is found to be inconclusive (or strong), no matter which of two priors on $m^{}_1$ is taken.
\end{abstract}


\def\thefootnote{\arabic{footnote}}
\setcounter{footnote}{0}
\newpage
\newpage
\section{Introduction}

The experimental observation of neutrinoless double-beta ($0\nu\beta\beta$) decays ${^{A}_Z}N \to {^A_{Z+2}}N + 2e^-$ of some heavy nuclei ${^{A}_Z}N$, which possess an even atomic number $Z$ and an even mass number $A$, is currently the most promising way to probe the Majorana nature of massive neutrinos and to prove the existence of lepton number violation in nature~\cite{Dolinski:2019nrj}. In the framework of three-flavor neutrino mixing, the $0\nu\beta\beta$ decays are mediated by three active neutrinos and the corresponding half-life of the $0\nu\beta\beta$-decaying even-even nuclear isotope is given by~\cite{Bilenky:2014uka}
\begin{eqnarray}\label{eq:halflife}
T^{0\nu}_{1/2} = G^{-1}_{0\nu} \cdot \left|{\cal M}^{}_{0\nu}\right|^{-2} \cdot \left|m^{}_{\beta\beta}\right|^{-2} \cdot m^2_e \; ,
\end{eqnarray}
where $G^{}_{0\nu}$ denotes the relevant phase-space factor, ${\cal M}^{}_{0\nu}$ is the nuclear matrix element (NME), and $m^{}_e = 0.511~{\rm MeV}$ is the electron mass.
As advocated by Particle Data Group~\cite{PDG2020}, the lepton flavor mixing matrix $U$ is usually parametrized in terms of three mixing angles $\{\theta^{}_{12}, \theta^{}_{13}, \theta^{}_{23}\}$, one Dirac-type CP-violating phase $\delta$ and two Majorana-type CP-violating phases $\{\phi^{}_{21}, \phi^{}_{31}\}$, i.e.,
\begin{eqnarray}\label{eq:pmns}
U = \left(
\begin{matrix}
c^{}_{12} c^{}_{13} & s^{}_{12} c^{}_{13} & s^{}_{13} e^{-{\rm i} \delta} \cr
-s^{}_{12} c^{}_{23} - c^{}_{12} s^{}_{13} s^{}_{23} e^{{\rm i} \delta} & c^{}_{12} c^{}_{23} -
s^{}_{12} s^{}_{13} s^{}_{23} e^{{\rm i} \delta} & c^{}_{13}
s^{}_{23} \cr
s^{}_{12} s^{}_{23} - c^{}_{12} s^{}_{13} c^{}_{23}
e^{{\rm i} \delta} & - c^{}_{12} s^{}_{23} - s^{}_{12} s^{}_{13}
c^{}_{23} e^{{\rm i} \delta} &   c^{}_{13} c^{}_{23} \cr
\end{matrix} \right) P^{}_{\nu}, \;
\end{eqnarray}
where $c^{}_{ij} \equiv \cos\theta^{}_{ij} $ and $ s^{}_{ij} \equiv \sin\theta^{}_{ij}$ (for $ij =12, 13, 23$) and $P_{\nu}^{} = \mathrm{diag}\{1, e^{{\rm i} \phi^{}_{21}/2},e^{{\rm i} \phi^{}_{31}/2}\}$.
The effective neutrino mass $|m^{}_{\beta \beta}|$ for $0\nu\beta\beta$ decays appearing in Eq.~(\ref{eq:halflife}) reads
\begin{eqnarray}\label{eq:mbb}
|m^{}_{\beta\beta}| \equiv \left|m^{}_1 \cos^2 \theta^{}_{13} \cos^2 \theta^{}_{12} e^{{\rm i}\rho} + m^{}_2 \cos^2 \theta^{}_{13} \sin^2 \theta^{}_{12} + m^{}_3 \sin^2 \theta^{}_{13} e^{{\rm i}\sigma}\right| \; ,
\end{eqnarray}
where $m^{}_i$ (for $i = 1, 2, 3$) stand for the absolute masses of three ordinary neutrinos. Out of three neutrino mixing angles only two $\{\theta^{}_{12}, \theta^{}_{13}\}$ are involved in the effective neutrino mass in Eq.~(\ref{eq:mbb}), where two Majorana-type CP-violating phases $\rho \equiv -\phi^{}_{21}$ and $\sigma \equiv (\phi^{}_{31} - \phi^{}_{21}) - 2\delta$ have been redefined as in Ref.~\cite{Xing:2014yka}. The other neutrino mixing angle $\theta^{}_{23}$ is irrelevant for $0\nu\beta\beta$ decays.

In the past few decades, neutrino oscillation experiments have measured  with a good precision the two neutrino mixing angles $\{\theta^{}_{12}, \theta^{}_{13}\}$, and two independent neutrino mass-squared differences $\Delta m^2_{21} \equiv m^2_2 - m^2_1$ and $|\Delta m^2_{31}| \equiv |m^2_3 - m^2_1|$~\cite{Wang:2015rma}. In the near future, the JUNO experiment will offer an unambiguous answer to whether neutrino mass ordering is normal $m^{}_1 < m^{}_2 < m^{}_3$ (NO) or inverted $m^{}_3 < m^{}_1 < m^{}_2$ (IO), and improve the precisions of all four parameters $\{\sin^2 \theta^{}_{12}, \sin^2 \theta^{}_{13}\}$ and $\{\Delta m^2_{21}, \Delta m^2_{31}\}$ to the ${\cal O}(0.1\%)$ level~\cite{Li:2013zyd, An:2015jdp, Vogel:2015wua, Cao:2017drk}. Given the precision data on these parameters, the observation of $0\nu\beta\beta$ decays will be extremely important in the determination of other fundamental parameters that cannot be probed in neutrino oscillation experiments, such as the absolute scale $m^{}_{\rm L}$ of neutrino masses, i.e., the lightest neutrino mass $m^{}_{\rm 1}$ (for NO) or $m^{}_3$ (for IO) and two Majorana CP phases $\{\rho, \sigma\}$. In particular, the experimental constraints on the Majorana CP phases can be obtained only in the lepton-number-violating processes, among which $0\nu\beta\beta$ decays should be most feasible and promising.

In this paper, we demonstrate that it is scientifically beneficial and even indispensable to reach the meV frontier of $|m^{}_{\beta\beta}|$, by quantitatively examining the projected sensitivities of future $0\nu\beta\beta$-decay experiments to the absolute neutrino masses and two Majorana CP phases. The main motivation for such an investigation is three-fold. First, the upper bound on the absolute scale of neutrino masses extracted from the tritium beta-decay experiments is $m^{}_\beta < 2.3~{\rm eV}$ at the $95\%$ confidence level (CL) from Mainz~\cite{Kraus:2004zw}, $m^{}_\beta < 2.2~{\rm eV}$ at the $95\%$ CL from Troitsk~\cite{Aseev:2011dq}, and $m^{}_\beta < 1.1~{\rm eV}$ at the $90\%$ CL from KATRIN~\cite{Aker:2019uuj}, where the effective neutrino mass $m^{}_\beta$ for beta decays is defined as $m^{}_\beta \equiv \left( m^2_1 |U^{}_{e1}|^2 + m^2_2 |U^{}_{e2}|^2 + m^2_3 |U^{}_{e3}|^2\right)^{1/2}$ with the moduli of the matrix elements of lepton flavor mixing matrix being $|U^{}_{e1}| = \cos \theta^{}_{13} \cos \theta^{}_{12}$, $|U^{}_{e2}| = \cos \theta^{}_{13} \sin \theta^{}_{12}$ and $|U^{}_{e3}| = \sin \theta^{}_{13}$ in the standard parametrization. The future operation of KATRIN~\cite{Osipowicz:2001sq, Wolf:2008hf} and
the next-generation tritium beta-decay experiment Project 8~\cite{Esfahani:2017dmu} will hopefully be able to bring the upper limit down to $m^{}_\beta \lesssim 0.2~{\rm eV}$ and $m^{}_\beta \lesssim 40~{\rm meV}$, respectively. On the other hand, the cosmological observations of cosmic microwave background by the Planck satellite gives the most restrictive bound on the sum of three neutrino masses $\Sigma \equiv m^{}_1 + m^{}_2 + m^{}_3 < 0.12~{\rm eV}$~\cite{Aghanim:2018eyx}. However, there is still a long way to go until the mass region of a few meV can be accessed. Second, if massive neutrinos are indeed Majorana particles, then two associated CP-violating phases $\{\rho, \sigma\}$ are fundamental parameters in nature and must be experimentally determined. At present, the $0\nu\beta\beta$ decays are the unique feasible pathway to get close to this goal~\cite{XZZ,Ge:2016tfx,XZ,XZ2,Penedo:2018kpc,Cao:2019hli,Ge:2019ldu}.
In this connection, the neutrino-antineutrino oscillations and other lepton-number-violating processes could in principle also provide some useful information about Majorana CP phases~\cite{Xing:2013ty, Xing:2013woa}, but the observations of these processes are currently still far away from reality. Even though a number of analytical studies of the effective neutrino mass $|m^{}_{\beta\beta}|$ have been performed in the literature~\cite{XZZ, Ge:2016tfx, XZ, XZ2, Penedo:2018kpc, Cao:2019hli, Ge:2019ldu}, some particular values of $|m^{}_{\beta\beta}|$ are assumed to derive the constraints on neutrino masses and Majorana CP phases. However, the effective neutrino mass $|m^{}_{\beta\beta}|$ itself is not a direct observable of $0\nu\beta\beta$-decay experiments. A robust statistical analysis is desirable to answer the following question: (i) given an experimental setup, what can we learn from a null signal after systematically taking into account the uncertainties of oscillation data, the nuclear matrix element and the phase-space factor? (ii) or conversely, to derive competitive bounds on the neutrino mass and Majorana phases, which kind of experimental setups will be required in the future?
Finally, the latest global-fit analysis of neutrino oscillation data yields a $2\sigma$ hint at the normal neutrino mass ordering~\cite{Esteban:2018azc}, so it is timely to investigate the physics potential of the future $0\nu\beta\beta$-decay experiments that aim at the ultimate discovery even in the NO case. Strategically speaking, whether the target value of the effective neutrino mass $|m^{}_{\beta\beta}|$ should be set to $10~{\rm meV}$ or $1~{\rm meV}$ makes a significant difference.

The remaining part of this paper is structured as follows. In Sec.~\ref{sec:exp}, the sensitivities of $0\nu\beta\beta$-decay experiments to the half-life $T_{1/2}^{0\nu}$ and to the effective neutrino mass $|m^{}_{\beta\beta}|$ are discussed. In Sec.~\ref{sec:massCP}, the sensitivities to the absolute neutrino mass scale and Majorana CP phases are examined by following the Bayesian statistics, where the physics potential of future experiments is investigated. Then we implement the Bayesian factors to discriminate between NO and IO, as well as the Dirac and Majorana nature of the massive neutrinos, in a quantitative way. Finally, we summarize our main conclusions in Sec.~\ref{sec:conc}.

\section{Sensitivities to $T^{0\nu}_{1/2}$ and $|m^{}_{\beta\beta}|$}
\label{sec:exp}
A number of nuclear isotopes have been found to be suitable for observing $0\nu\beta\beta$ decays~\cite{Dolinski:2019nrj, Bilenky:2014uka}. In the present work we take the nuclear isotope ${^{136}}{\rm Xe}$ for illustration, which has been implemented in the currently leading $0\nu\beta\beta$-decay experiments (e.g., KamLAND-Zen~\cite{KamLAND-Zen:2016pfg} and EXO-200~\cite{Auger:2012ar,Albert:2014awa}), and the other candidates can be studied in a similar way.

It should be helpful to first establish the relation between an experimental configuration and its sensitivity to the effective neutrino mass $|m^{}_{\beta\beta}|$. This task has already been accomplished in Ref.~\cite{Agostini:2017jim}, but we shall reproduce the main results in this section for completeness and for setting up our notations for later discussions. As is well known, for a given setup of the $0\nu\beta\beta$-decay experiment, its sensitivity to the half-life $T^{0\nu}_{1/2}$ for $0\nu\beta\beta$ decays can be derived by using the following formula~\cite{Agostini:2017jim}
\begin{eqnarray}\label{eq:shl}
T^{0\nu}_{1/2} =\ln{2} \cdot \frac{N^{}_{\rm A} \cdot \xi \cdot \epsilon}{m^{}_{\rm iso} \cdot S^{}_{}(B)} \; ,
\end{eqnarray}
where $N^{}_{\rm A}=6.022 \times 10^{23}$ is the Avogadro's constant, $m^{}_{\rm iso}$ is the molar mass of the relevant nuclear isotope, $\xi \equiv M^{}_{\rm iso}\cdot t$ is the total exposure of the experiment with $M^{}_{\rm iso}$ being the total target mass of the decaying isotope and $t$ being the running time of the experiment, and $\epsilon$ is the detection efficiency of the signal event. In addition, $S^{}_{}(B)$ in Eq.~(\ref{eq:shl}) is defined as the expected number of signal events within the region of interest (ROI) when a specified fraction $q$ of a set of identical experiments can report a discovery of the $0\nu\beta\beta$ decay signal at the CL $\geq p$~\cite{Punzi:2003bu}, where the dependence on the total number of background events $B \equiv b \cdot \xi$ has been explicitly stated with $b$ being the background index (in units of counts per ton$\cdot$yr).

Given the expectation value $\mu$ of the total number of events, the number of counts $n$ truly observed in the experiment statistically fluctuates according to the Poisson distribution, for which the probability distribution function (PDF) is given by ${\rm PDF}^{}_{\rm Poisson} (n, \mu) = e^{-\mu} \mu^n / {n!}$ and the corresponding cumulative distribution function (CDF) reads
\begin{eqnarray}\label{eq:cdf}
{\rm CDF}^{}_{\rm Poisson} (n, \mu) & \equiv & \sum^n_{k = 0} {\rm PDF}^{}_{\rm Poisson} (k, \mu) \;.
\end{eqnarray}
The expectation value of the signal event number to set the experimental sensitivity $S^{}_{}(B)$ can be figured out by solving the equations
\begin{eqnarray}\label{eq:ses}
{\rm CDF}^{}_{\rm Poisson} (n^{}_{p }, B) \geq p   \; , \quad
\overline{\rm CDF}^{}_{\rm Poisson} (n^{}_{p }, B + S)=  q  \; ,
\end{eqnarray}
where $n^{}_{p }$ is the smallest number of counts to exclude the null-signal hypothesis at the CL $\geq p$, and $\overline{\rm CDF}^{}_{\rm Poisson} = 1 - {\rm CDF}^{}_{\rm Poisson}$ is the complementary function of the CDF. In the extreme background-free case, any positive signal events mean a discovery, i.e. the required number of counts is always $n^{}_{p} = 1$ regardless of $p$, and $q$ can be interpreted as the probability that an experiment can report a positive signal (otherwise null signal). The median sensitivity usually adopted in the literature refers to a discovery probability of $50\%$, which is quite reasonable when the background is large and events are Gaussian distributed. In the background-free scenario, however, this implies that there is a probability of $50\%$ to observe null signal; therefore a larger value of $q$ (e.g. $90\%$, $95\%$, etc) should be adopted such that the obtained sensitivity is more solid and close to the true case.

\begin{figure}[t!]
	\begin{center}
		\subfigure{
			\includegraphics[width=0.47\textwidth]{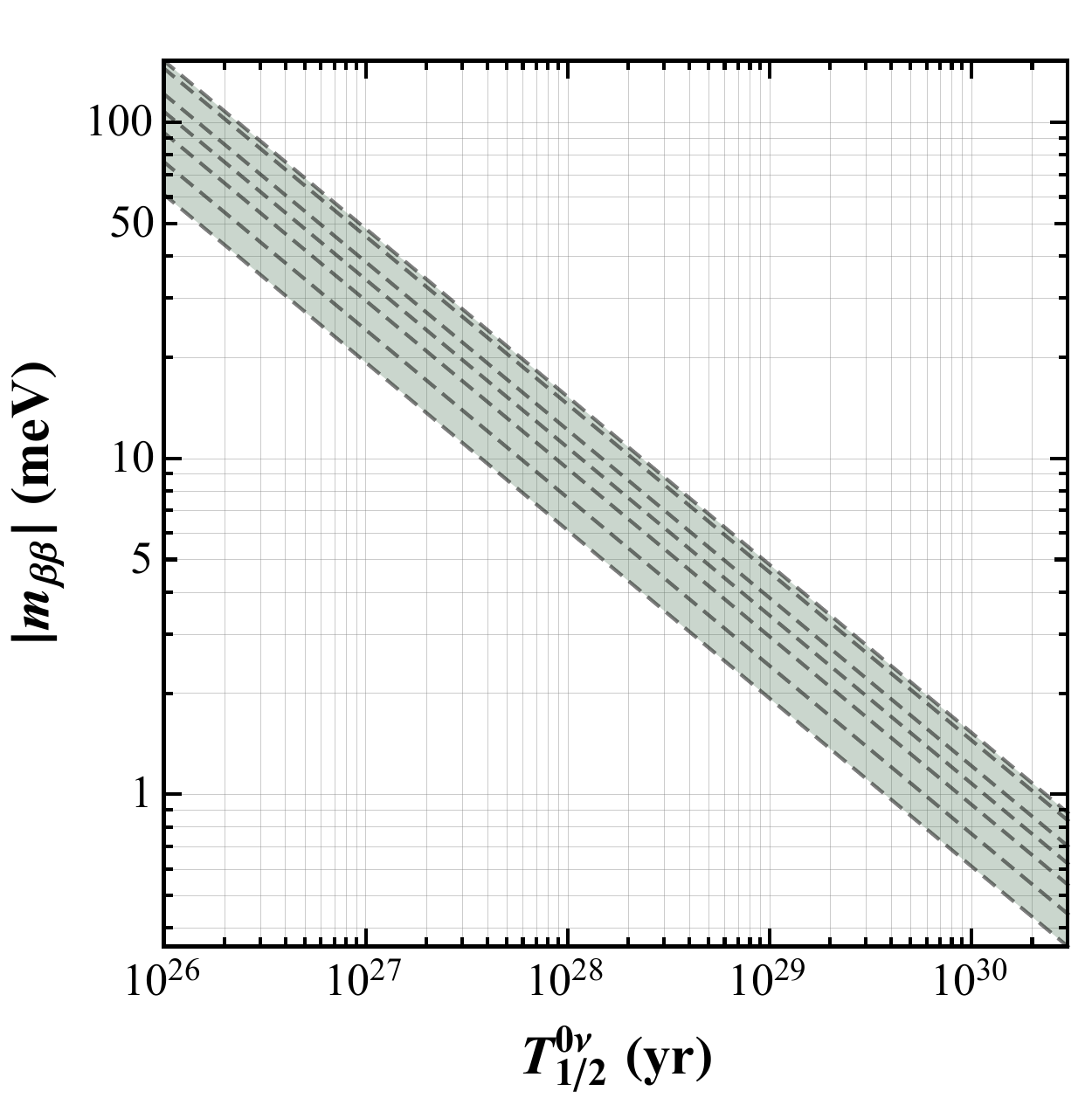} }
		\subfigure{
			\includegraphics[width=0.48\textwidth]{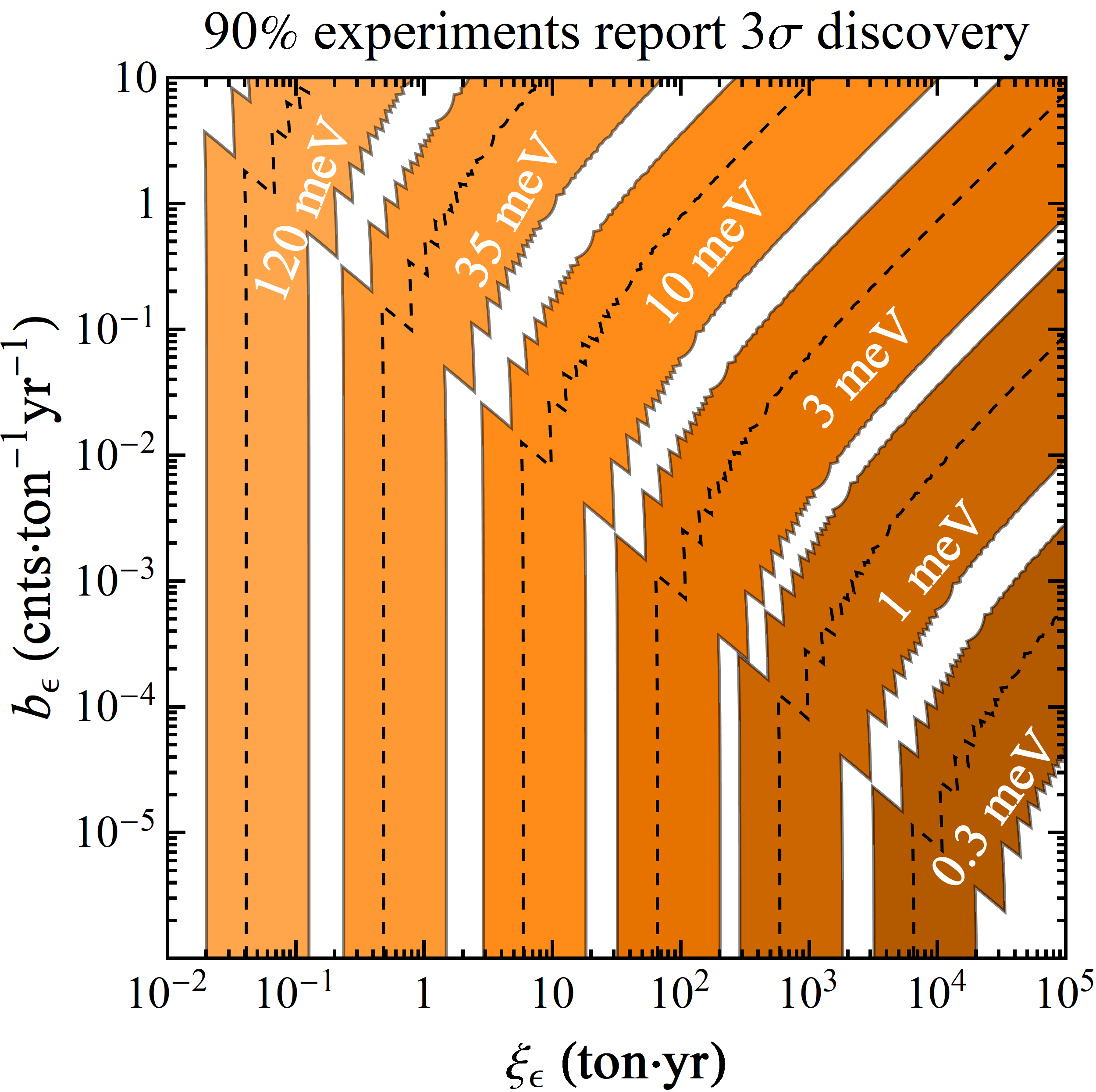} }
	\end{center}
	\vspace{-0.3cm}
	\caption{The relationship between the $0\nu\beta\beta$-decay half-life $T^{0\nu}_{1/2}$ of $^{136}{\rm Xe}$ and the effective neutrino mass $|m^{}_{\beta\beta}|$ ({\it left panel}), where the phase-space factor $G^{}_{0\nu} = 3.79\times 10^{-14}~{\rm yr}^{-1}$ and the values of NME of $1.68 \lesssim |{\cal M}^{}_{0\nu}| \lesssim 4.20$ as compiled in Ref.~\cite{Caldwell:2017mqu} have been considered. The  sensitivity to $T^{0\nu}_{1/2}$ and $|m^{}_{\beta\beta}|$ at the $3\sigma$ CL for $90\%$ experiments has been plotted as contours in the plane of the effective exposure $\xi^{}_{\epsilon}$ and the effective background index $b^{}_{\epsilon}$ ({\it right panel}). The colored bands in both left and right panels stem from the NME uncertainty. The corresponding dashed curves in the middle of the band of the right panel is obtained by using the average NME value of $|{\cal M}^{}_{0\nu}| = 2.94$.}
	\label{fig:E_B_sensitivity}
\end{figure}

With the above definitions in mind, we are ready to derive the sensitivities to $T^{0\nu}_{1/2}$ and $|m^{}_{\beta\beta}|$ for any given experimental setup of the exposure and background index. For later convenience, we introduce the effective exposure $\xi^{}_{\epsilon} \equiv \xi \cdot \epsilon$ and the effective background index $b^{}_{\epsilon} \equiv b / \epsilon$ such that the signal detection efficiency $\epsilon$ in Eq.~(\ref{eq:shl}) is no longer present explicitly. In the left panel of Fig.~\ref{fig:E_B_sensitivity}, the relationship between $T^{0\nu}_{1/2}$ and $|m^{}_{\beta\beta}|$, as indicated in Eq.~(\ref{eq:halflife}), has been shown for the nuclear isotope ${^{136}}{\rm Xe}$. In the numerical calculations, the phase-space factor $G^{}_{0\nu} = 3.79\times 10^{-14}~{\rm yr}^{-1}$ with the axial vector coupling $g^{}_{\rm A} = 1.27$ has been used~\cite{Suhonen:1998ck, Rodejohann:2011mu,Kotila:2012zza}, while the NME for the $0\nu\beta\beta$ decays of ${^{136}}{\rm Xe}$ has been taken from Table II of Ref.~\cite{Caldwell:2017mqu}, where one can find that the theoretical predictions for NME via different methods span a wide range of $1.68 \lesssim |{\cal M}^{}_{0\nu}| \lesssim 4.20$. The gray band in the left panel of Fig.~\ref{fig:E_B_sensitivity} shows the NME uncertainties, and seven typical values of NME have been plotted as dashed lines. Therefore, depending on the NME, $|m^{}_{\beta\beta}| = 10~{\rm meV}$ and $|m^{}_{\beta\beta}| = 1~{\rm meV}$ correspond to the half-life of $T^{0\nu}_{1/2} \in \left(3.8\times 10^{27} \cdots 2.5\times 10^{28}\right)~{\rm yr}$ and $T^{0\nu}_{1/2} \in \left(3.8\times 10^{29} \cdots 2.5\times 10^{30}\right)~{\rm yr}$, respectively. This observation can be perfectly understood by noting that the half-life $T^{0\nu}_{1/2}$ is inversely proportional to $|m^{}_{\beta\beta}|^2$ as in Eq.~(\ref{eq:halflife}).

The experimental sensitivity to $|m^{}_{\beta\beta}|$ should be given together with specific values of $p$ and $q$ defined in Eq.~(\ref{eq:ses}). For instance, the sensitivity at the $3\sigma$ CL for $90\%$ experiments corresponds to $p =99.73\%$ and $q =90\%$. In the right panel of Fig.~\ref{fig:E_B_sensitivity}, we have displayed the contours of the sensitivity to $|m^{}_{\beta\beta}|$ at the $3\sigma$ CL for $90\%$ identical experiments in the plane of the effective exposure $\xi^{}_\epsilon$ and the effective background index $b^{}_\epsilon$. Note that the discrete characteristic of the Poisson distribution becomes apparent when the background counts reach the threshold value of $\mathcal{O}(1)$. Similar results can also be found in Fig.~3 of Ref.~\cite{Agostini:2017jim} with a different concerned region of $\xi^{}_\epsilon$ and $b^{}_\epsilon$ and a smoothed Poisson distribution. The dashed curves denote the contours of the sensitivity to $|m^{}_{\beta\beta}|$, where the white colored number in the band is calculated with the average NME value $|{\cal M}^{}_{0\nu}| = 2.94$. Each colored band stands for the same sensitivity to $|m^{}_{\beta\beta}|$ as indicated, and its  width signifies the NME uncertainty. It should be noticed that to achieve the sensitivity of $|m^{}_{\beta\beta}| = 1~{\rm meV}$ is very challenging. Even with a background-free environment (namely, $b^{}_\epsilon \rightarrow 0$), an effective exposure of at least $\xi^{}_\epsilon \simeq 300~{\rm ton \cdot yr}$ is required to reach the sensitivity of $|m^{}_{\beta\beta}| = 1~{\rm meV}$ at the $3\sigma$ CL.

\section{Sensitivity to fundamental parameters}\label{sec:appa}
\label{sec:massCP}

\subsection{The Bayesian approach}

The Bayesian statistics provides us a logical and practical approach to comparing different models as well as inferring the posterior probability distributions of model parameters. According to the Bayesian theorem, the posterior probability of a hypothesis in light of the experimental data $\mathcal{D}$ is
\begin{eqnarray}
{P}(\mathcal{H}^{}_{i} | \mathcal{D}) =\frac{P(\mathcal{D}|\mathcal{H}^{}_{i}) P(\mathcal{H}^{}_{i})}{P(\mathcal{D})}\;,
\end{eqnarray}
where $\mathcal{H}^{}_{i}$ stands for the hypothesis with $i$ being the index of different models, and $P(\mathcal{H}^{}_{i})$ is the prior probability for the model to be true. In addition, $P(\mathcal{D}|\mathcal{H}^{}_{i})$ is identical to the so-called evidence $\mathcal{Z}^{}_{i}$, which is the total likelihood to observe $\mathcal{D}$ given the hypothesis $\mathcal{H}^{}_{i}$, and ${P}(\mathcal{D}) = \sum^{}_{i}P(\mathcal{D}|\mathcal{H}^{}_{i}) P(\mathcal{H}^{}_{i})$ can be regarded as a normalization factor that fixes $\sum^{}_{i} {P}(\mathcal{H}^{}_{i} | \mathcal{D}) = 1$. The model favored by the experimental data among a set of models can be selected by taking their posterior ratios, i.e.,
\begin{eqnarray} \label{eq:odds}
\frac{P(\mathcal{H}^{}_{i} | \mathcal{D})}{P(\mathcal{H}^{}_{j} | \mathcal{D})} = \frac{\mathcal{Z}^{}_{i}}{\mathcal{Z}^{}_{j}} \frac{P(\mathcal{H}^{}_{i})}{P(\mathcal{H}^{}_{j})}\;.
\end{eqnarray}
If we assume no prior preference for any models, the Bayes factor $\mathcal{B} \equiv \mathcal{Z}^{}_{i}/\mathcal{Z}^{}_{j}$ can directly reflect the odds of different models. We will adopt the Jeffreys scale \cite{Trotta:2008qt} to interpret the Bayes factor.

The posteriors in the parameter space of a specific model can also be updated in light of the experimental data. The posterior probability distribution of the model parameter set $\Theta$ can be derived according to
\begin{eqnarray}
{P}(\Theta | \mathcal{H}^{}_{i},\mathcal{D}) =\frac{P(\mathcal{D}|\mathcal{H}^{}_{i}, \Theta) P(\Theta|\mathcal{H}^{}_{i})}{P(\mathcal{D} | \mathcal{H}^{}_{i})}\;,
\end{eqnarray}
where $P(\mathcal{D}|\mathcal{H}^{}_{i}, \Theta)$ denotes the likelihood function in the assumption that the model $\mathcal{H}^{}_{i}$ with the parameter set $\Theta$ is true, and $P(\Theta|\mathcal{H}^{}_{i})$ is the prior probability of $\Theta$. Here $P(\mathcal{D} | \mathcal{H}^{}_{i})$ is the aforementioned evidence $\mathcal{Z}^{}_{i}$, which can be obtained by integrating over all model parameters,
\begin{eqnarray} \label{eq:BayesFactor}
\mathcal{Z}^{}_{i} = \int P(\mathcal{D}|\mathcal{H}^{}_{i}, \Theta) P(\Theta|\mathcal{H}^{}_{i}) \mathrm{d} \Theta\;.
\end{eqnarray}
We will use the \texttt{MultiNest} routine for the Bayesian analysis~\cite{Feroz:2007kg,Feroz:2008xx,Feroz:2013hea}.

\subsection{Sensitivities to $m^{}_1$, $\rho$ and $\sigma$}

The next-generation $0\nu\beta\beta$-decay experiments aim to cover entirely the whole range of $|m^{}_{\beta\beta}|$ in the IO case. The lower boundary of $|m^{}_{\beta\beta}|$ is always lying above $10~{\rm meV}$, which will be taken as a representative value for the sensitivity of next-generation experiments to $|m^{}_{\beta\beta}|$.
The target value of $|m^{}_{\beta\beta}|\sim 10~{\rm meV}$ can be hopefully achieved in a number of proposed experiments, e.g., LEGEND~\cite{Abgrall:2017syy}, CUPID~\cite{Wang:2015taa}, nEXO~\cite{Albert:2017hjq}, JUNO Xe-LS~\cite{Zhao:2016brs} and PandaX-III~\cite{Chen:2016qcd}.
Moreover, we try to explore the physics potential of the $0\nu\beta\beta$-decay experiment with a sensitivity to $|m^{}_{\beta\beta}| \simeq 1~{\rm meV}$ in the NO case. Thus two scenarios will be considered:
\begin{itemize}
\item {\bf Setup-I} with the total exposure $\xi = 50~{\rm ton} \cdot 5~{\rm yr}$, the background index $b = 1.35~{\rm ton^{-1} \cdot yr^{-1}}$, and the detection efficiency $\epsilon = 0.634$. Such a setup is inspired by the preliminary study of the future JUNO Xe-LS experiment in Ref.~\cite{Zhao:2016brs}. Given this experimental setup, one can derive its projected sensitivity to half-life $T^{0\nu}_{1/2} = 6.24 \times 10^{27}~{\rm yr}$ at the $3\sigma$ CL, which can be translated into the sensitivity to the effective neutrino mass $|m^{}_{\beta\beta}| = \left(7.9\cdots 19.7\right)~{\rm meV}$ (depending on the NME) at the same CL.

\item {\bf Setup-II} with the total exposure $\xi = 400~{\rm ton} \cdot 5~{\rm yr}$, the background index $b = 0~{\rm ton^{-1} \cdot yr^{-1}}$, and the detection efficiency $\epsilon = 1$. In comparison with the previous setup, the exposure is now increased by one order of magnitude, while the background is assumed to be vanishing. As it is very challenging in reality to achieve these improvements, this experimental setup may just stand for the ultimate goal of the $0\nu\beta\beta$-decay experiments in the far future. With this ideal setup, we find that the sensitivity to $T^{0\nu}_{1/2}$ is $2.67 \times 10^{30}~{\rm yr}$ at the $3\sigma$ CL, or equivalently the sensitivity to $|m^{}_{\beta\beta}|$ is $(0.38 \cdots 0.95)~{\rm meV}$.
\end{itemize}
For each specific experimental setup, one is able to examine its sensitivities to the fundamental parameters, such as the lightest neutrino mass $m^{}_1$ and two Majorana CP phases $\{\rho, \sigma\}$, which is the main task in this subsection.
The Bayesian approach will be implemented to derive the posterior distributions of the model parameters and to select favorable models~\cite{Trotta:2008qt, Zhang:2015kaa}. The strategy for our statistical analysis is outlined as below.

First, we assume that the future experiments would have not discovered any signals of $0\nu\beta\beta$ decays, so the observed events should be ascribed solely to the background. For the background event number $B$ and a hypothetical signal event number $N^{}_{0\nu}$, the probability to observe the number $n^{}_{\rm tot}$ of total events in the ROI is determined by the likelihood function
\begin{eqnarray} \label{eq:lhJUNO}
\mathcal{L}^{\rm meV}_{0\nu\beta\beta}(N^{0\nu}) = \frac{(N^{0\nu}+B)^{n^{}_{\rm tot}}}{n^{}_{\rm tot}!} \cdot e^{-(N^{0\nu}+B)} \; ,
\end{eqnarray}
where the Poisson distribution is adopted.

Second, the prior distributions of two relevant neutrino mixing angles $\{\sin^2 \theta^{}_{12}, \sin^2\theta^{}_{13}\}$ and two neutrino mass-squared differences $\{\Delta m^2_{21}, \Delta m^2_{31}\}$ are taken to be flat in some ranges, which are chosen to be wide enough to cover the latest global-fit results of all neutrino oscillation data in Ref.~\cite{Esteban:2018azc}. The unconstrained Majorana CP phases $\{\rho, \sigma\}$ are uniformly distributed in the whole range $[0, 360^\circ)$. As a fundamental parameter, the lightest neutrino mass $m^{}_1/{\rm eV}$ or its logarithm $\log^{}_{10}(m^{}_1/{\rm eV})$ will be uniformly distributed in the range of $m^{}_1/{\rm eV} \in [10^{-7}, 10]$ or $\log^{}_{10}(m^{}_1/{\rm eV}) \in [-7, 1]$, which will be referred to as the flat or log prior on $m^{}_1$ in the following discussions.
Note that one may also adopt the flat or log prior on the sum of three neutrino masses $\Sigma$ instead of $m^{}_{1}$. We have numerically checked that with a sensitivity of $|m^{}_{\beta\beta}| \lesssim 10~{\rm meV}$, these two prior options of $\Sigma$ lead to posteriors very similar to that with a flat prior on $m^{}_{1}$.
In connecting the fundamental parameters to the hypothetical signal events in $0\nu\beta\beta$-decay experiments, one must specify the phase-space factor $G^{}_{0\nu}$ and the NME value $|{\cal M}^{}_{0\nu}|$. In our calculations for $^{136}{\rm Xe}$, the phase-space factor $G^{}_{0\nu}$ is supposed to be Gaussian distributed with the central value and $1\sigma$ error as found in Ref.~\cite{Kotila:2012zza},
namely $G^{}_{0\nu} = 3.79\times 10^{-14}~{\rm yr}^{-1}$ with an error of $0.1\%$,
while the NME $|{\cal M}^{}_{0\nu}|$ is uniformly distributed in the range $[1.68, 4.20]$ as obtained in various nuclear models~\cite{Caldwell:2017mqu}. Now, all the priors of model parameters in our analysis have been specified.

Third, the posterior distributions can be derived by imposing the experimental likelihood information of both the existing data and the simulated data of future $0\nu\beta\beta$-decay experiments. To be explicit, the likelihood functions of neutrino oscillation parameters are extracted from the global-fit analysis of Ref.~\cite{Esteban:2018azc}. For the likelihood of the future $0\nu\beta\beta$-decay experiments, we will generate the Asimov data, for which the simulated event number is the same as the theoretical expectation. In general, one should follow the Feldman-Cousins approach~\cite{Feldman:1997qc} by taking the median projection of Monte Carlo simulations. In the assumption of null signals, we can simply set $n^{}_{\rm tot} = B$ in Eq.~(\ref{eq:lhJUNO}). For each set of parameters in the model under test, one can predict the expected number of events $N^{0\nu}_{}$ as explained in the second step and find out its associated likelihood by using Eq.~(\ref{eq:lhJUNO}). It is worthwhile to mention that one may generate the data based on a true event signal, and the estimation of model parameters in this case can also be studied.

To demonstrate the independent constraining power of future $0\nu\beta\beta$-decay experiments on the absolute scale of neutrinos masses, we do not include the likelihood of other existing non-oscillation experiments in limiting the neutrino masses and Majorana CP phases. But they will be included for the discriminations between NO and IO as well as the Majorana and Dirac nature of neutrinos in Sec.~\ref{sec:MDS}. The likelihood information about the effective neutrino mass $m^{}_\beta$ in beta decays, the effective neutrino mass $|m^{}_{\beta\beta}|$ in $0\nu\beta\beta$ decays and the sum of three neutrino masses $\Sigma \equiv m^{}_1 + m^{}_2 + m^{}_3$ is extracted from the existing beta-decay experiments (i.e., Mainz~\cite{Kraus:2004zw}, Troitsk~\cite{Aseev:2011dq} and KATRIN~\cite{Aker:2019uuj}), $0\nu\beta\beta$-decay experiments (including GERDA~\cite{Agostini:2019hzm}, KamLAND-Zen~\cite{KamLAND-Zen:2016pfg}, EXO~\cite{Albert:2014awa} and CUORE~\cite{Alduino:2017ehq}) and the cosmological observations~\cite{Aghanim:2018eyx}, respectively.

\begin{figure}[t!]
	\begin{center}
		\hspace{-0.6cm}
		\includegraphics[width=0.52\textwidth]{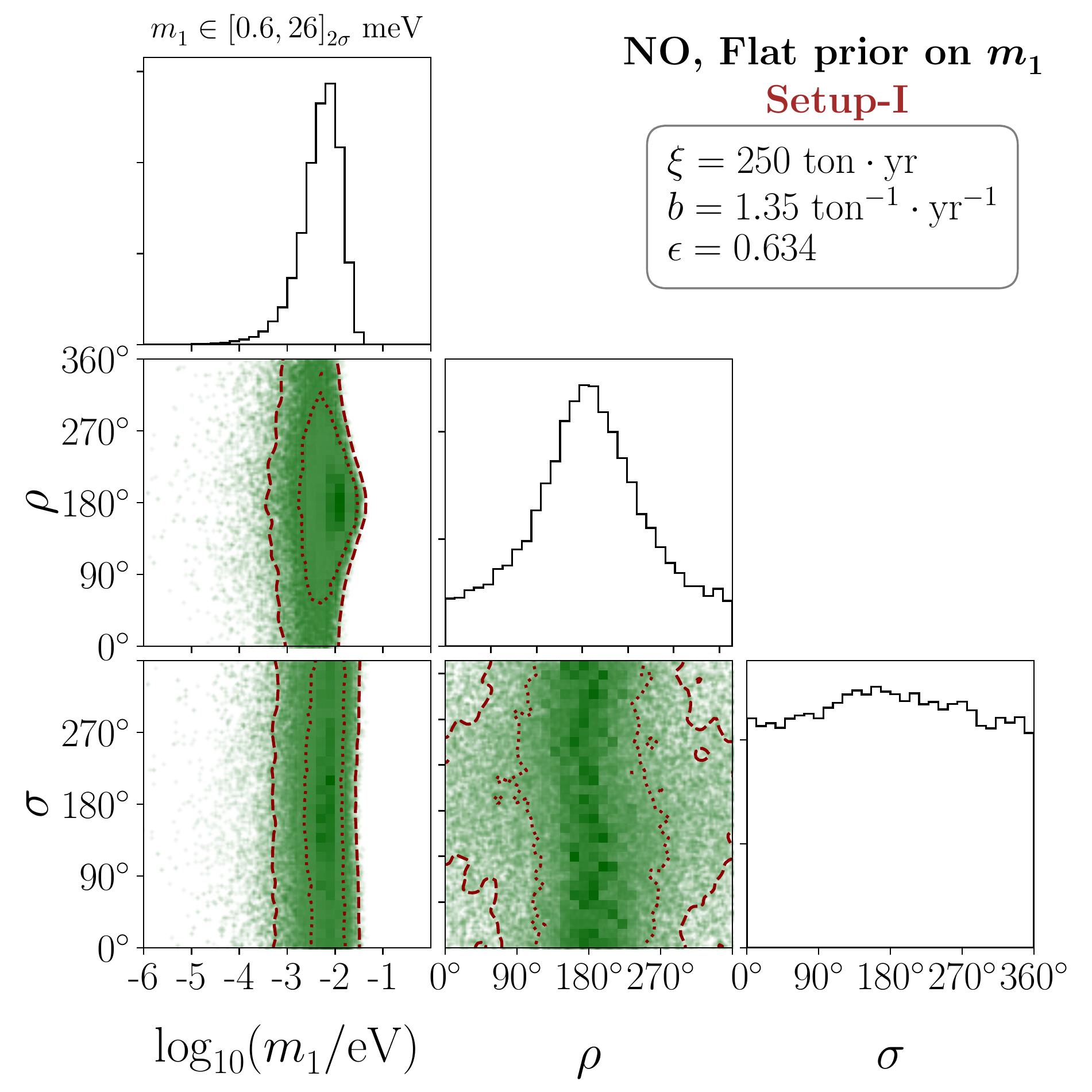}
		\hspace{-0.5cm}
		\includegraphics[width=0.52\textwidth]{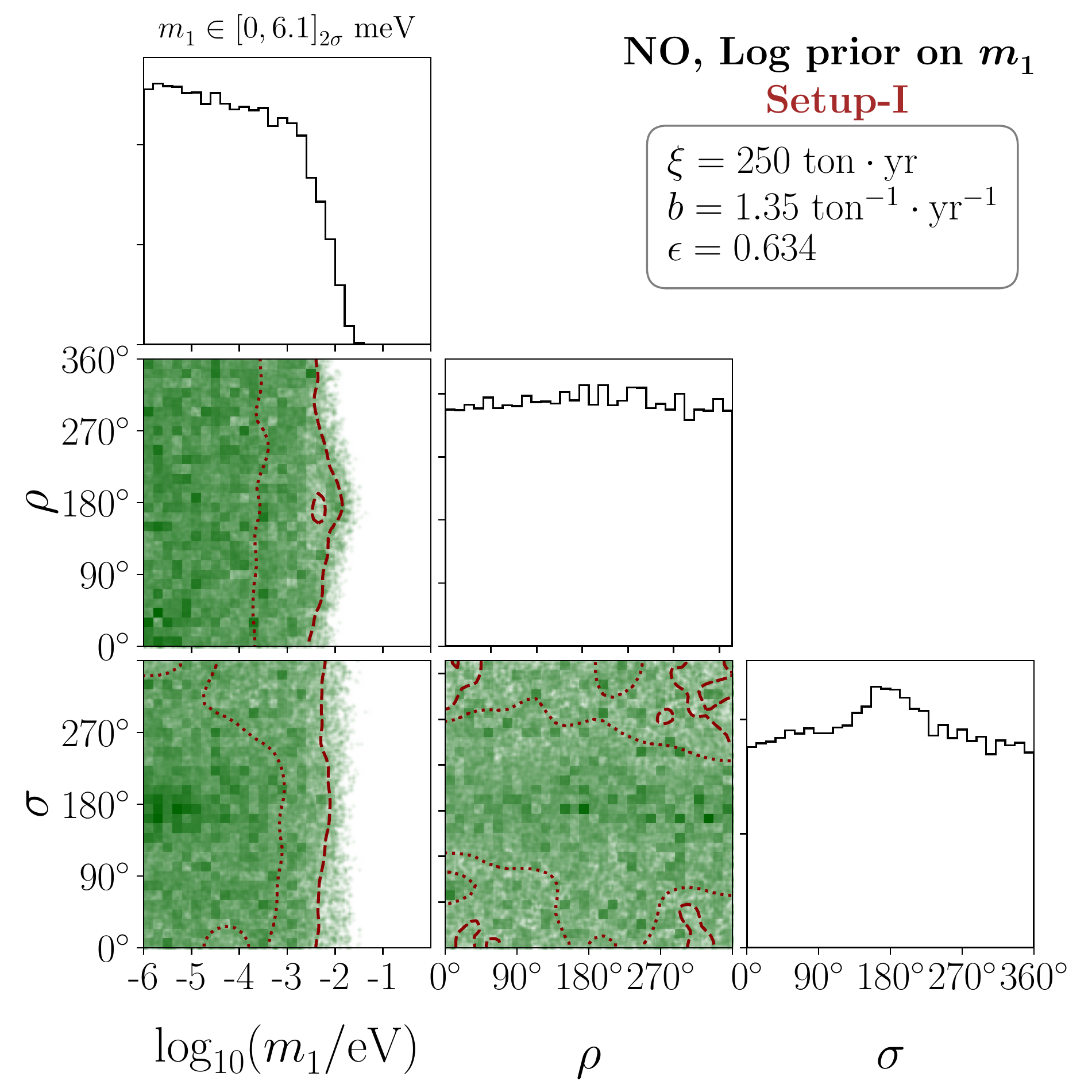}
		\\[0ex]
		\hspace{-0.6cm}
		\includegraphics[width=0.52\textwidth]{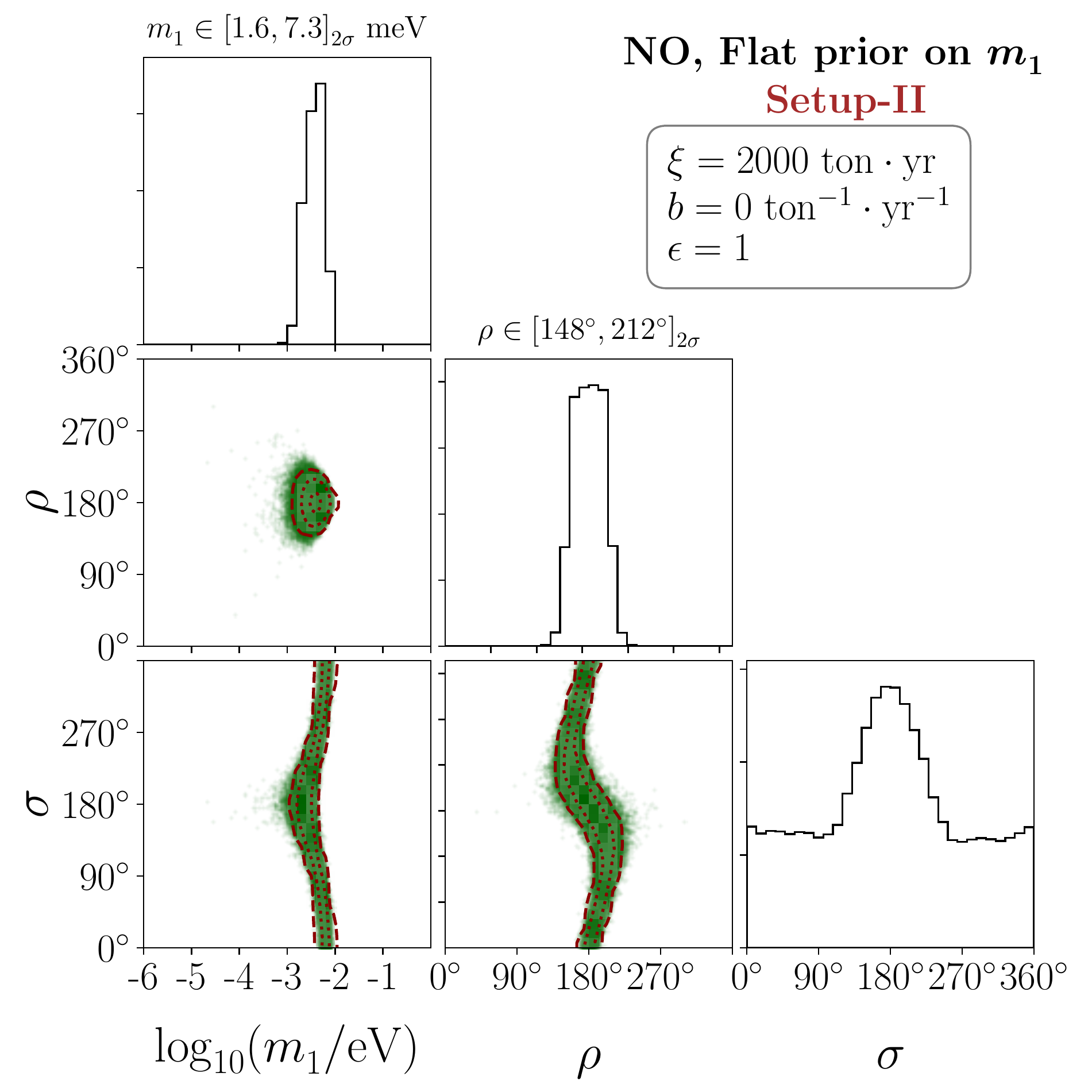}
		\hspace{-0.5cm}
		\includegraphics[width=0.52\textwidth]{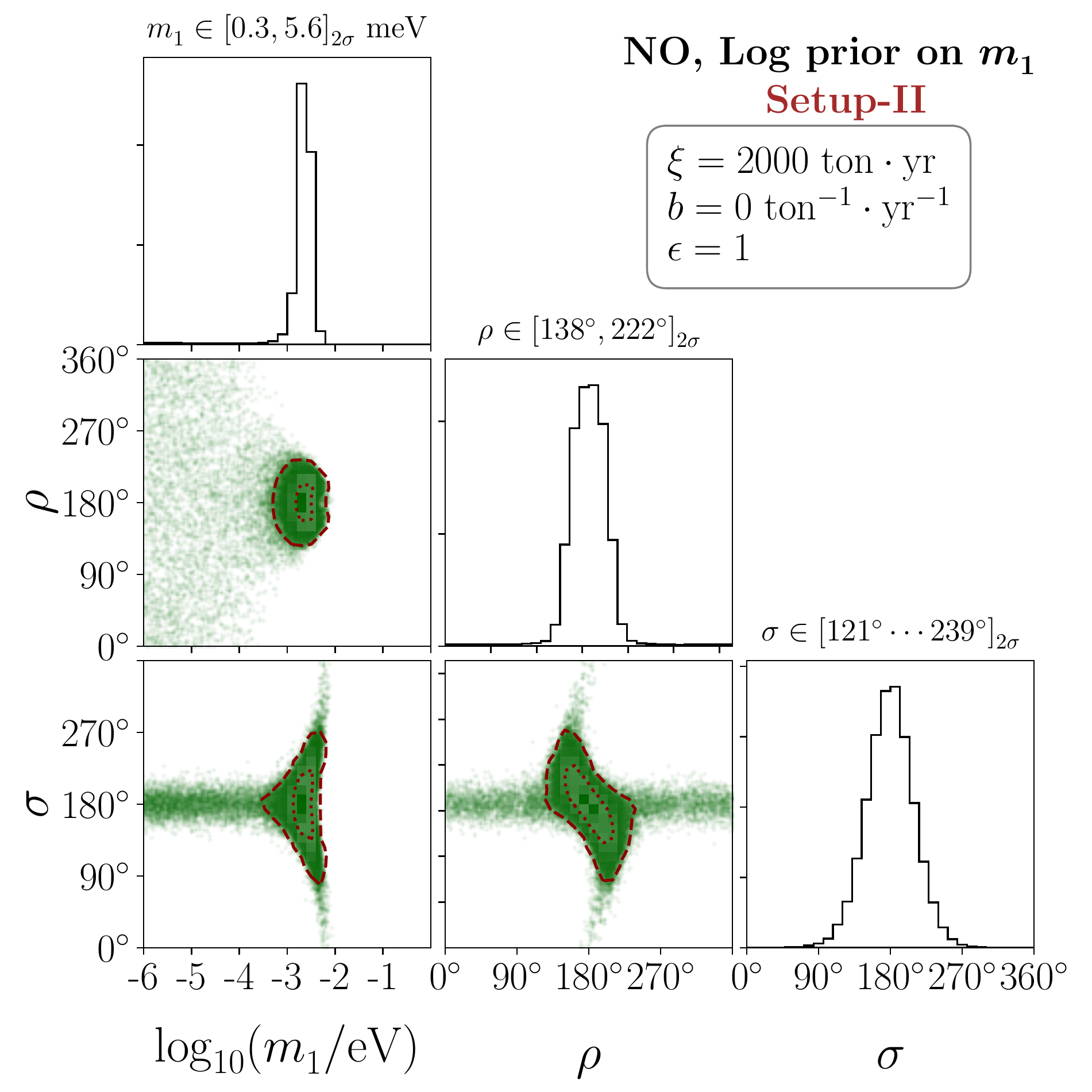}
	\end{center}
	\vspace{-0.3cm}
	\caption{The posterior distributions of the lightest neutrino mass $m^{}_{\rm 1}$, and the Majorana CP phases $\rho$ and $\sigma$ in assumption of the null signal in future $0\nu\beta\beta$-decay experiments. Their individual posteriors as well as their correlations have been shown. The regions formed by the green points stand for the correlations of these three parameters. The dashed (dotted) red contours surround the $2\sigma$ ($1\sigma$) regions of highest posterior densities. The histograms stand for the posteriors of each individual parameters of $m^{}_{\rm 1}$, $\rho$ and $\sigma$ with proper normalizations.}
	\label{fig:posterior_setup}
\end{figure}

Assuming a null signal in the aforementioned experimental setup of $0\nu\beta\beta$ decays and following the strategy outline above, one can set limits on the absolute scale of neutrino masses as well as the Majorana CP phases. In Fig.~\ref{fig:posterior_setup}, the posterior distributions of $m^{}_{\rm 1}$, $\rho$ and $\sigma$ have been presented.
The upper (lower) two panels stand for the cases of {\bf Setup-I} ({\bf Setup-II}) with the flat prior on $m^{}_{1}$ and the log prior on $m^{}_{1}$, respectively. In each panel, the green regions of points demonstrate the correlations of $m^{}_{1}$, $\rho$ and $\sigma$ in their posterior distributions, while the dashed (dotted) red contours surround the $2\sigma$ ($1\sigma$) regions of the highest posterior densities (HPD). HPD means that the posterior densities are the same along the contours. The individual posterior distribution of $m^{}_{1}$, $\rho$ or $\sigma$ is obtained as the black histogram, with a title above signifying its corresponding $95\%$ credible interval whenever it is significant. The quoted credible intervals of $m^{}_{1}$ are obtained by treating $\log^{}_{10}(m^{}_{1}/{\rm eV})$ as the model parameter. By definition the credible interval may change if one rescales the model parameter. Some comments on the numerical results are helpful.
\begin{itemize}
\item With {\bf Setup-I} corresponding to the sensitivity of $|m^{}_{\beta\beta}| \approx 10~{\rm meV}$, the null signal of $0\nu\beta\beta$ decays can constrain the lightest neutrino mass $m^{}_{1}$ into the $95\%$ credible range
    \begin{eqnarray}\label{eq:setup1m1}
    \left\{
      \begin{array}{lll}
        m^{}_1 \in [0.6 \cdots 26]~{\rm meV} \; , &~& \hbox{for flat prior on $m^{}_1$ ;} \\
        m^{}_1 \in [0 \cdots 6.1]~{\rm meV} \; , &~& \hbox{for log prior on $m^{}_1$ .}
      \end{array}
    \right.
    \end{eqnarray}
For the flat prior, the interval of $m^{}_{1}$ is bounded from below because of the prior effect.
For the log prior, the upper limit of the credible interval is slightly subject to the ad hoc lower bound when we set the prior. For instance, if we shift this prior bound from $10^{-7}~{\rm eV}$ to $10^{-4}~{\rm eV}$, the upper limit will be changed from $6.1~{\rm meV}$ to $9.2~{\rm meV}$ accordingly.
The upper bound can be transformed into the limit on the sum of neutrino masses $\Sigma$ by using the best-fit values of mass-squared differences ~\cite{Esteban:2018azc} as
    \begin{eqnarray}\label{eq:setup1Sigma}
    \left\{
      \begin{array}{lll}
        \Sigma \equiv m^{}_1 + m^{}_2 + m^{}_3 < 0.11~{\rm eV} \;, &~& \hbox{for flat prior on $m^{}_1$ ;} \\
        \Sigma \equiv m^{}_1 + m^{}_2 + m^{}_3 < 0.067~{\rm eV} \;, &~& \hbox{for log prior on $m^{}_1$ ,}
      \end{array}
    \right.
    \end{eqnarray}
which are very competitive with the cosmological bounds of Planck. We should emphasize that the $0\nu\beta\beta$-decay experiments can provide a direct information on the absolute scale of the lightest neutrino mass instead of a bound on the sum of all neutrino masses $\Sigma$ as in cosmology. The limits on $m^{}_{1}$ from future cosmological surveys are not expected to be so strong by transforming from the future cosmological bounds on $\Sigma$, e.g., from $\Sigma \lesssim 0.087~{\rm eV}$ to $m^{}_{1} \lesssim 16~{\rm meV}$ at $95\%$ CL\cite{Dvorkin:2019jgs}. The null-signal constraints on the Majorana CP phases $\rho$ and $\sigma$ are rather weak. The $95\%$ credible interval of the strongest one reads $\rho \in [28^{\circ} \cdots 342^{\circ}]$ when we adopt the flat prior on $m^{}_{1}$, while the other Majorana CP phase $\sigma$ is almost unconstrained.
\item With {\bf Setup-II} corresponding to the sensitivity of $|m^{}_{\beta\beta}| \approx 1~{\rm meV}$, the null-signal simulation will exclude a large fraction of the parameter space of $m^{}_{1}$, $\rho$ and $\sigma$. As shown in the second row of Fig.~\ref{fig:posterior_setup}, very informative conclusions can be made in this case. The lightest neutrino mass $m^{}_{1}$ can be constrained into the $95\%$ credible range
    \begin{eqnarray}\label{eq:setup2m1}
    \left\{
      \begin{array}{lll}
        m^{}_1 \in [1.6 \cdots 7.3]~{\rm meV} \; , &~& \hbox{for flat prior on $m^{}_1$ ;} \\
        m^{}_1 \in [0.3 \cdots 5.6]~{\rm meV} \; , &~& \hbox{for log prior on $m^{}_1$ ,}
      \end{array}
    \right.
    \end{eqnarray}
which are much better than other observational constraints from beta decays and cosmology in the foreseeable future. These two limits are mostly stable against a change on the model parameter from $\log^{}_{10}(m^{}_{1}/{\rm eV})$ to $m^{}_{1}/{\rm eV}$ in obtaining the credible intervals.
Apparently, the lower bounds on $m^{}_1$ in Eq.~(\ref{eq:setup2m1}) arise from the ``well"-like structure of $|m^{}_{\beta\beta}|$~\cite{XZ, XZ2}. In addition, the constraints on the Majorana CP phases at the $95\%$ CL turn out to be
    \begin{eqnarray}\label{eq:setup2phase}
    \left\{
      \begin{array}{lll}
       \rho \in [148^\circ \cdots 212^\circ] \; , &~& \hbox{for flat prior on $m^{}_1$ ;} \\
        \rho \in [138^\circ \cdots 222^\circ]\; , &~& \hbox{for log prior on $m^{}_1$ ,}
      \end{array}
    \right.
    \end{eqnarray}
The constraint on $\rho$ is almost independent of the priors on $m^{}_1$, which contains only $20\%$ of the whole range of $[0\cdots 360^\circ)$. The limits on $\sigma$ are not so strong for both priors, e.g.
$\sigma \in [121^\circ \cdots 239^\circ]$ for the log prior on $m^{}_{1}$ and basically unconstrained for the flat prior. The correlations among $m^{}_{1}$, $\rho$ and $\sigma$ agree well with the analytical results in Ref.~\cite{Cao:2019hli}.
\end{itemize}

\begin{figure}[t!]
	\begin{center}
		\subfigure{
			\hspace{-0.2cm}
			\includegraphics[width=0.48\textwidth]{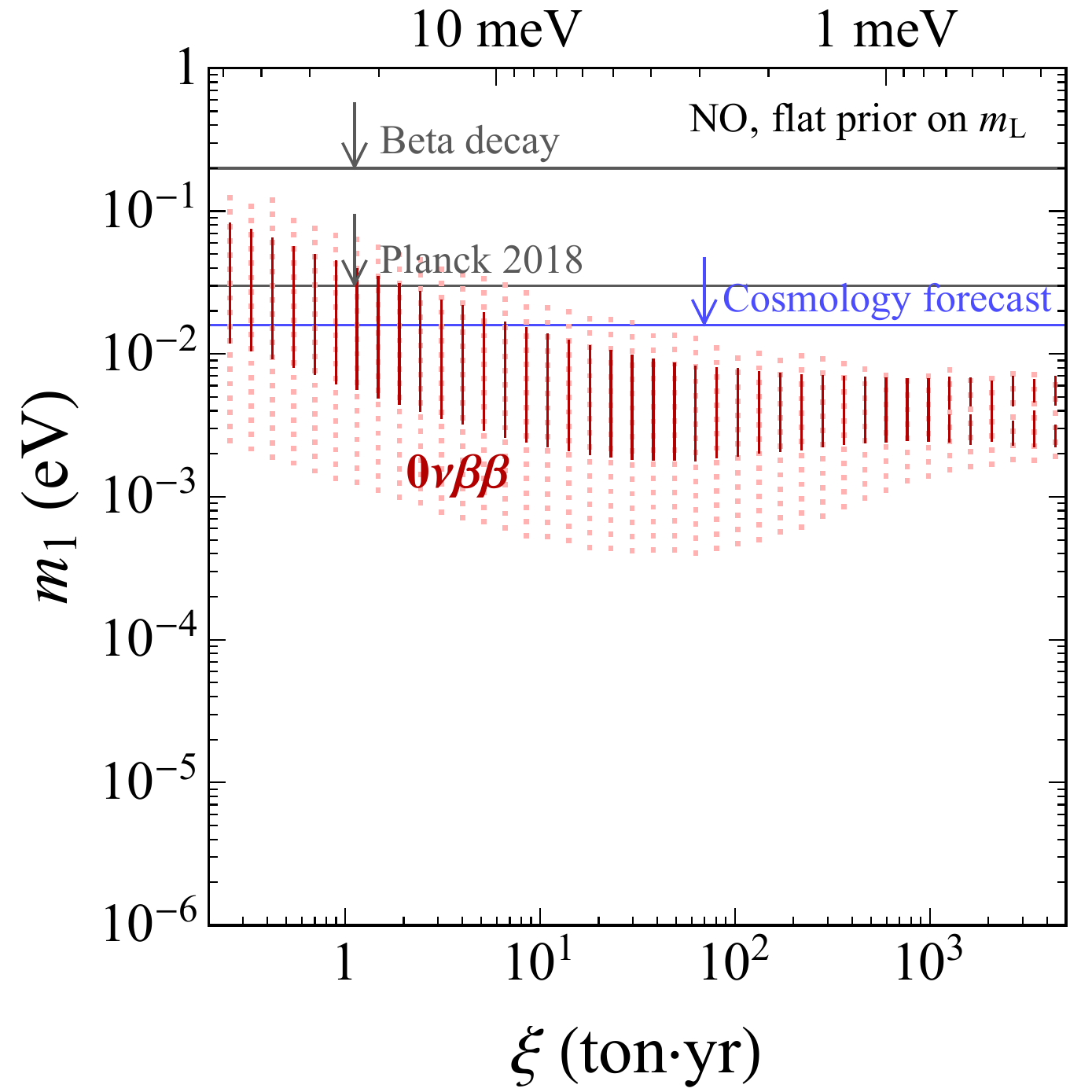} }
		\subfigure{
			\includegraphics[width=0.48\textwidth]{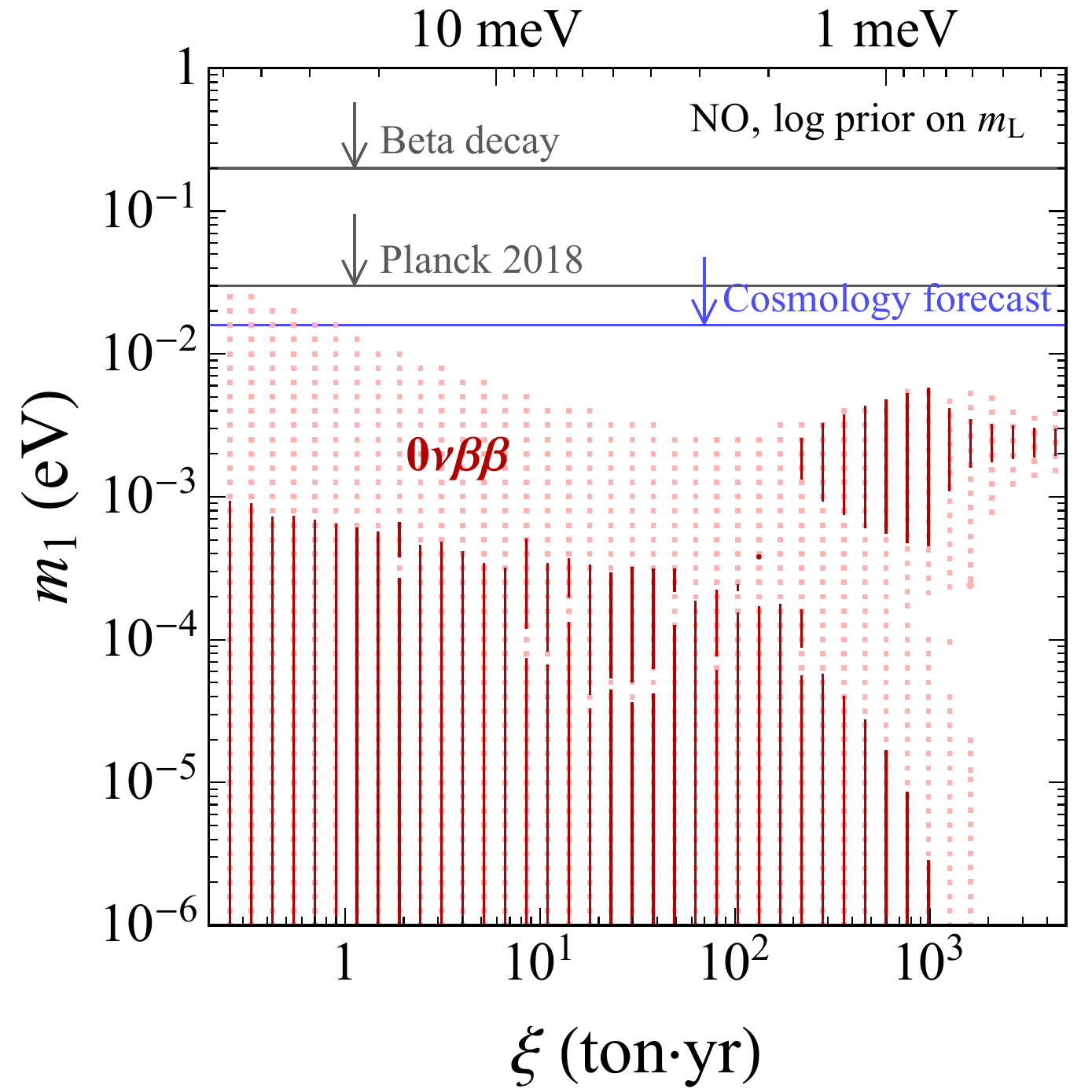} }
	\end{center}
	\vspace{-0.3cm}
	\caption{The $68\%$ (solid red lines) and $95\%$ (dotted red lines) credible intervals of $m^{}_{\rm 1}$ for different values of the exposure $\xi$ with a vanishing background and a full efficiency $\epsilon = 1$. Null signal in future $0\nu\beta\beta$-decay experiments is assumed. The top horizontal axis signifies the corresponding $3\sigma$ sensitivities to $|m^{}_{\beta\beta}|$ with the average NME of $|{\cal M}^{}_{0\nu}| = 2.94$ for ${^{136}}{\rm Xe}$..}
	\label{fig:E_varying}
\end{figure}
With those two specific setups, we have shown their constraining power on the lightest neutrino mass $m^{}_{1}$. In more general cases, we vary the exposure $\xi$ with the background-free assumption, and plot the $1\sigma$ (solid red lines) and $2\sigma$ (dotted red lines) credible intervals of $m^{}_{1}$ under the null-signal assumption in Fig.~\ref{fig:E_varying}.
We can notice an apparent converging behavior for two priors. This observation makes sense for the Bayesian analysis, namely, as more and more data have been collected the impact of priors will eventually fade away. For the case of the log prior on $m^{}_{1}$, a lower bound on $m^{}_{1}$ appears only after the $3\sigma$ sensitivity of the setup to $|m^{}_{\beta\beta}|$ has reached around $1~{\rm meV}$. This result is quite meaningful, as there is only a very small fraction of the parameter space in the ``well"-like structure. See, e.g., blue curves of Fig.~3 in Ref.~\cite{Huang:2019qvq}. One can obtain a lower limit on $m^{}_{1}$ only when the parameter space with $m^{}_{1} \rightarrow 0~{\rm meV}$ is highly disfavored, which requires a sensitivity of $|m^{}_{\beta\beta}| \lesssim 1~{\rm meV}$.
In the right panel of Fig.~\ref{fig:E_varying}, as the exposure increases, the credible intervals do not strictly shrink for the exposure between $10^{2}~{\rm ton\cdot yr}$ and $10^{3}~{\rm ton\cdot yr}$. This effect is due to the shift of probability from $m^{}_{1} \lesssim 10^{-3}~{\rm eV}$ towards the range $10^{-3}~{\rm eV} \lesssim m^{}_{1} \lesssim 10^{-2}~{\rm eV}$ in the well-like structure. Beyond a critical value of the exposure, the credible intervals will gradually stop changing as the well-like structure spans a certain range.
The bounds from the KATRIN projection and Planck 2018 results are transformed into those on $m^{}_{1}$ and shown as gray horizontal lines for comparison, while the future cosmology sensitivity to $m^{}_{1}$ at  $95\%$ CL corresponding to $\sigma(\Sigma) \sim 14~{\rm meV}$ ~\cite{Dvorkin:2019jgs} is given as the horizontal blue line. One can clearly note the advantage of $0\nu\beta\beta$-decay experiments in probing the absolute scale of neutrino masses when the $\mathcal{O}({\rm meV})$ sensitivity is achieved~\cite{Cao:2019hli}.

\begin{figure}[t!]
	\begin{center}
			\includegraphics[width=0.48\textwidth]{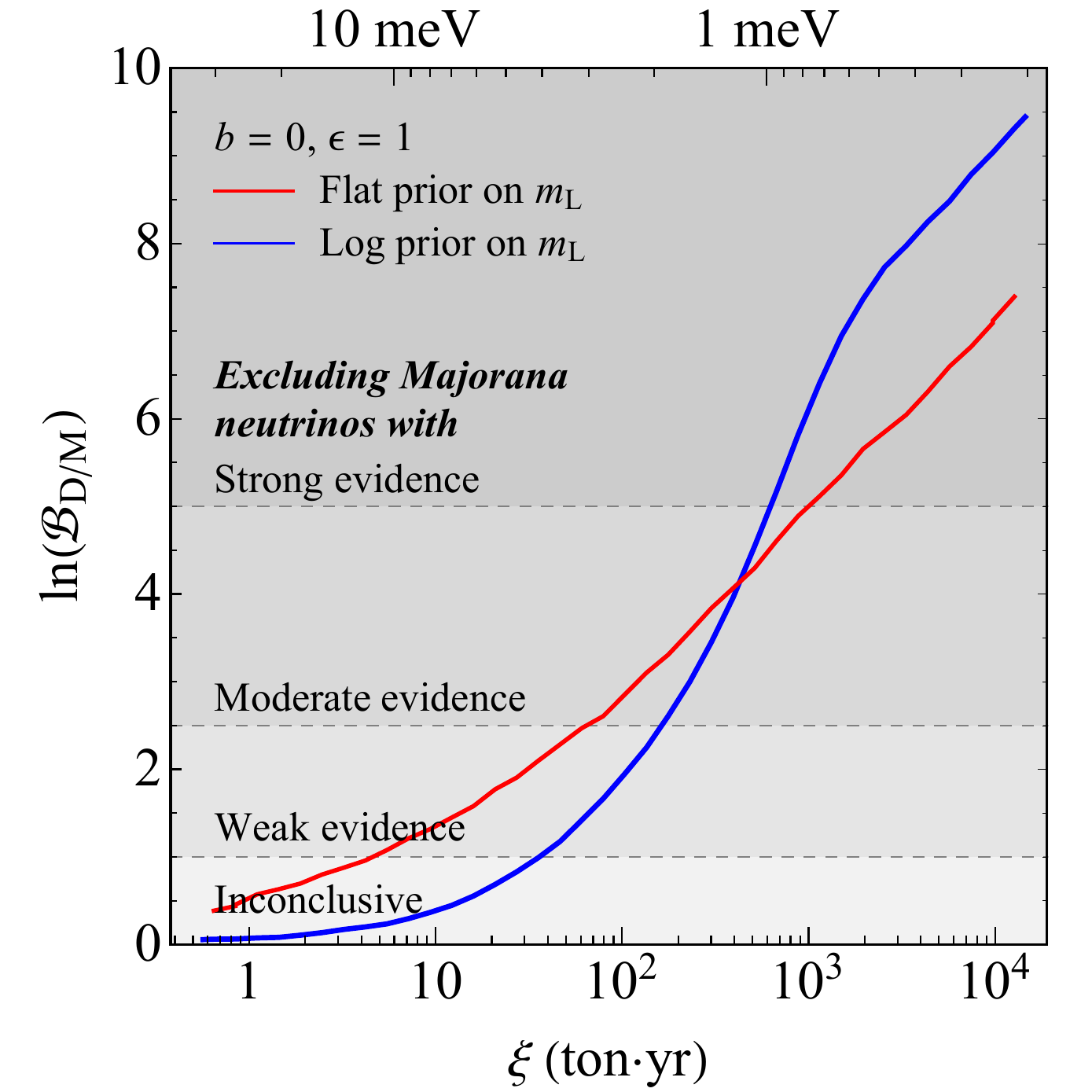}
	\end{center}
	\vspace{-0.3cm}
	\caption{
	The Bayesian factor $\ln(\mathcal{B}^{}_{\rm D/M})$, which reflects the odds of Dirac neutrinos over Majorana ones, for different values of the exposure $\xi$ with a vanishing background and a full efficiency $\epsilon = 1$. Null signal in future $0\nu\beta\beta$-decay experiments is assumed. The top horizontal axis signifies the corresponding $3\sigma$ sensitivities to $|m^{}_{\beta\beta}|$ with the average NME of $|{\cal M}^{}_{0\nu}| = 2.94$ for ${^{136}}{\rm Xe}$.}
	\label{fig:BDM}
\end{figure}
\subsection{NO vs. IO and Majorana vs. Dirac}\label{sec:MDS}

Although it is reasonable to assume that the neutrino mass ordering is normal, as indicated by the latest global-fit analysis of neutrino oscillation data, and massive neutrinos are Majorana particles, as in a class of seesaw models of neutrino masses, we can use the Bayesian approach to perform a model comparison. Such a study is based on no prior preference for NO or Majorana neutrinos and maximizes the information from current and future experimental observations.

In the Bayesian analysis the preference of NO over IO can be represented by the Bayes factor $\mathcal{B}^{}_{\rm N/I}$ which is defined as the ratio of the evidence of NO to that of IO~\cite{Zhang:2015kaa}.
In either case of NO or IO, the fundamental parameters are the same, including two neutrino mixing angles, two neutrino mass-squared differences, two Majorana CP phases and the lightest neutrino mass. Given these parameters, one can compute the effective mass $|m^{}_{\beta\beta}|$, which together with the phase-space factor and the NME will predict the $0\nu\beta\beta$-decay rate. Then the likelihoods for the signal events can be calculated for a nominal $0\nu\beta\beta$-decay experiment. Finally, the Bayes factor can be conveniently obtained by integrating the product of priors and likelihoods over the model parameters as in Eq.~(\ref{eq:BayesFactor}).
Before taking account of the simulated likelihood of future $0\nu\beta\beta$-decay experiments, one can already notice some preference for NO from the cosmological observations and from the existing $0\nu\beta\beta$-decay searches. With the log prior on $m^{}_{\rm L}$ (i.e., $m^{}_{\rm 1}$ for NO and $m^{}_{\rm 3}$ for IO), one finds $\ln(\mathcal{B}^{}_{\rm N/I}) = 1.08$, while with the flat prior on $m^{}_{\rm L}$, one finds a similar result $\ln(\mathcal{B}^{}_{\rm N/I}) = 1.25$, implying a weak evidence of NO according to the Jeffreys scale \cite{Trotta:2008qt}. This conclusion has been reached without including the NO preference from neutrino oscillation experiments. By taking account of the null-signal simulation of {\bf Setup-I}, the Bayes factor increases to $\ln(\mathcal{B}^{}_{\rm N/I}) = 12.5$ for the log prior on $m^{}_{\rm L}$ and $\ln(\mathcal{B}^{}_{\rm N/I}) = 12.1$ for the flat prior on $m^{}_{\rm L}$. The statistical odds of NO over IO is very large, i.e, a factor of $\mathcal{B}^{}_{\rm N/I} \sim 10^{5}$. This result indicates a super strong discriminating power for neutrino mass ordering in future $0\nu\beta\beta$-decay experiments with a $|m^{}_{\beta\beta}| \simeq 10~{\rm meV}$ sensitivity like {\bf Setup-I}.
	
The hypotheses of Majorana and Dirac neutrinos can also be tested by the Bayesian analysis. For the Dirac scenario, one can simply take the half-life of $0\nu\beta\beta$ decays to be infinitely long in generating the posterior distributions. The statistical odds of Dirac over Majorana neutrinos can be measured by the Bayes factor $\mathcal{B}^{}_{\rm D/M}$. After including the  null-signal likelihood of {\bf Setup-I}, the logarithm of the Bayes factor for NO reads $\ln(\mathcal{B}^{}_{\rm D/M}) = 0.16$ with the log prior on $m^{}_{\rm 1}$ and $\ln(\mathcal{B}^{}_{\rm D/M}) = 1.0$ with the flat prior on $m^{}_{\rm 1}$. The statistical odds is not yet adequate to infer a moderate evidence, i.e., $\ln(\mathcal{B} ) = 2.5$, since a considerable fraction of the Majorana parameter space is not covered with {\bf Setup-I}. However, in the IO case, one can find $\ln(\mathcal{B}^{}_{\rm D/M}) = 11.7$ and $\ln(\mathcal{B}^{}_{\rm D/M}) = 12.3$ for the log and flat priors on $m^{}_{3}$, respectively. As has been expected, almost all parameter space of the IO case for Majorana neutrinos is covered with {\bf Setup-I}. If the neutrino mass ordering is found to be IO in future neutrino oscillation experiments like JUNO, one can discriminate the Dirac hypothesis from the Majorana one with a very high statistical significance in the JUNO Xe-LS experiment of $0\nu\beta\beta$ decays or other similar experiments with a competitive sensitivity.
For the NO case, to have an adequate evidence in favor of the Dirac hypothesis over the Majorana one, one must go beyond {\bf Setup-I}. We show in Fig.~\ref{fig:BDM} the Bayesian factor $\ln(\mathcal{B}^{}_{\rm D/M})$ as a function of the exposure with the background-free assumption. It can be clearly observed that to obtain a strong evidence, i.e., $\ln(\mathcal{B} ) = 5$, the exposure should be as large as $10^3~{\rm ton \cdot yr}$ while keeping the background vanishing, which corresponds to the ultimate meV sensitivity of the $0\nu\beta\beta$-decay experiment.
In other words, if there is null $0\nu\beta\beta$-decay signal at the meV frontier, we can then claim that the Majorana nature of neutrinos is excluded with a strong evidence.
However, it is worthwhile to stress that the conclusions here are based on the standard mechanism of exchanging three light neutrinos, which may not apply to the $0\nu\beta\beta$ decays induced by some non-standard physics (e.g., sterile neutrinos and left-right symmetric models~\cite{Bilenky:2014uka, Rodejohann:2011mu}).

\section{Summary}\label{sec:conc}

In order to explore the physics potential of future $0\nu\beta\beta$-decay experiments with a sensitivity of $|m^{}_{\beta\beta}| \approx 1~{\rm meV}$, we have investigated the projected constraints on the lightest neutrino mass $m^{}_1$ and the Majorana CP phases, in the assumption of a null signal. For comparison, the experimental setup for the sensitivity of $|m^{}_{\beta\beta}| \approx 10~{\rm meV}$ is also considered. Our main results and conclusions are summarized in Eqs.~(\ref{eq:setup1m1})-(\ref{eq:setup2phase}), where the Bayesian approach is adopted for statistical analysis.

We believe that our analysis is very important and suggestive for setting up the future program for $0\nu\beta\beta$-decay experiments. As already pointed out in Ref.~\cite{Cao:2019hli}, if the experimental sensitivity of $|m^{}_{\beta\beta}| = 1~{\rm meV}$ is ultimately realized, the determination of absolute neutrino masses and the constraints on Majorana CP phases are very promising, which cannot be reached in other types of future neutrino experiments. We have examined these issues in a quantitative way by performing the Bayesian analysis. Furthermore, the determination of neutrino mass ordering and the Majorana or Dirac nature of massive neutrinos are also studied. Certainly, to achieve all these goals, one has to make great efforts in increasing the target mass and reducing the background by two orders of magnitude compared to the present design of next-generation $0\nu\beta\beta$-decay experiments. These technical challenges will be left for more future works~\cite{Giuliani:2019uno}.

\section*{Acknowledgements}

The authors are indebted to Profs. Jun Cao, Yu-Feng Li, Manfred Lindner, Yi-Fang Wang, Liang-Jian Wen, Zhi-zhong Xing and Zhen-hua Zhao for helpful discussions. This work was supported in part by the Alexander von Humboldt Foundation, by the National Key R\&D Program of China under Grant No. 2018YFA0404100, by the National Natural Science Foundation of China under Grant No.~11775232 and No.~11835013, and by the CAS Center for Excellence in Particle Physics.

\end{document}